# 基于多层特征表示的光滑调制神经网络下车辆 24 色长尾识别

胡明娣[1)]，白龙[1)]，李英[1)]，赵思蕊[2)]，陈恩红[2)]

[1)](西安邮电大学 通信与信息工程学院，西安 710121)

[2)](中国科技技术大学 计算机科学与技术学院，合肥 230026)

**摘 要** 车辆颜色识别在智能交通管理和刑侦辅助判案中起着重要作用，而当前存在的车辆颜色识别研究最多涉及 13 类颜色且识别准确率较低，难以满足实际应用。为此，本文自建了一个包括 24 类车辆颜色的基准数据集（Vehicle Color-24），包括了从 100 个小时的城市道路监控视频里面截取的 10091 张车辆图片。此外，为解决 Vehicle Color-24 数据集存在的长尾分布和现有方法识别率低的问题，本文提出了基于多层特征表示的光滑调制神经网络（Smooth Modulated Neural Network with Multi-layer Feature Representation, SMNN-MFR）用于 24 类车辆颜色识别。SMNN-MFR 包括特征提取、多尺度特征融合、建议框生成和光滑调制四个部分，并在 Vehicle Color-24 基准数据集上训练和验证了模型的有效性。综合实验表明，本文算法在 24 类颜色基准数据库的平均识别准确率为 94.96%，比 Faster RCNN 网络提高了 33.47%。此外，模型识别 8 类颜色时的平均准确率为 97.25%，比同类数据库的算法检测精度有所提高。与此同时，通过可视化和消融实验也证明了我们网络设置的合理性和各模块的有效性。代码和数据库公布在：https://github.com/mendy-2013。

# Vehicle 24-Color Long Tail Recognition Based on Smooth Modulation Neural Network with Multi-layer Feature Representation

HU Ming-Di[1)], BAI Long[1)], LI Ying[1)], ZHAO Si-Rui[2)], CHEN En-Hong[2)]

[1)](Department of Communication and Information Engineering, Xi’an University of Posts & Telecommunications, Xi’an 710121)

[2)](Department of Computer Science and Technology, University of Science and Technology of China, Hefei 230026)

**Abstract** Vehicle color recognition plays an important role in intelligent traffic management and criminal investigation assistance. However, the current vehicle color recognition research involves at most 13 types of colors and the recognition accuracy is low, which is difficult to meet practical applications. To this end, this paper has built a benchmark dataset (Vehicle Color-24) that includes 24 types of vehicle colors, including 10091 vehicle pictures taken from 100 hours of urban road surveillance videos. In addition, in order to solve the problem of long tail distribution in Vehicle Color-24 dataset and low recognition rate of existing methods, this paper proposes a Smooth Modulated Neural Network with Multi-layer Feature Representation (SMNN-MFR) is used for 24 types of vehicle color recognition. SMNN-MFR includes four parts: feature extraction, multi-scale

收稿日期：2021年7月13日；最终修改稿收到日期：2021年-月-日. 本课题得到陕西省重点研发计划项目(2018KW050)资助. **胡明娣(通信作者)**，女，1970年生，博士，副教授，中国计算机学会(CCF)会员(17401M)，主要研究领域为图像识别、目标检索与分类、机器学习、人工智能、模糊信息处理.E-mail: humingdi2005@xupt.edu.cn. **白龙**，男，1997年生，硕士，主要研究领域为机器学习、深度神经网络、图像目标识别.E-mail: 1901210100@stu.xupt.edu.cn. **李英**，女，1997年生，硕士，主要研究领域为机器学习、深度神经网络、图像目标识别.E-mail: 1901200009@stu.xupt.edu.cn. **赵思蕊**，男，1991年生，博士，中国计算机学会(CCF)会员(B6323M)，主要研究领域为计算机视觉、深度学习、小样本学习. E-mail: sirui@mail.ustc.edu.cn. **陈恩红**，男，1968年生，博士，教授，中国计算机学会(CCF)会员(09201F)，主要研究领域为机器学习、数据挖掘、社交网络. E-mail: cheneh@ustc.edu.cn.
第1作者手机号码: 180 8926 6816, E-mail: humingdi2005@xupt.edu.cn.

feature fusion, suggestion frame generation and smooth modulation. The model is trained and verified on the Vehicle Color-24 benchmark dataset. Comprehensive experiments show that the average recognition accuracy of the algorithm in the 24 categories of color benchmark databases is 94.96%, which is 33.47% higher than the Faster RCNN network. In addition, the average accuracy rate of the model when recognizing 8 types of colors is 97.25%, and the detection accuracy of algorithms in similar databases is improved. At the same time, visualization and ablation experiments also proved the rationality of our network settings and the effectiveness of each module. The code and database are published at: https://github.com/mendy-2013.

**Key words** Target detection; Smooth modulation neural network; Multi-layer feature representation; Long tail recognition; Vehicle color-24 benchmark dataset

## 1 引言

截至 2020 年底，全国机动车保有量达 3.72 亿辆，车辆管理面临前所未有的压力。一方面，智能交通系统（Intelligent Transportation System，ITS）普遍存在管理质量差、效率低的问题；另一方面，以车辆为犯罪工具的刑侦案件越来越多，公安机关追踪、识别嫌疑车辆的效率仍有待提升。目前车辆管理主要是依据车牌、颜色、车型等属性来实时决策[1]，如果车辆更换、篡改、遮挡、拆卸或伪造车牌时，摄像头不能实时捕捉车牌，便不能依靠车牌快速识别车辆做出实时决策，给侦查办案增加难度。而车辆颜色作为获取车辆信息的另一个重要属性[2]，可以辅助交通执法，提高车辆身份识别的可靠性。

由于涂料和调色的复杂，车辆颜色丰富且个性，至少有上百种颜色。现有的车辆颜色识别研究[3-5]最多涉及 13 类车辆颜色的识别，包括 Chen 等人[3]建立的车辆颜色数据集，含白色、黑色、灰色、红色、青色、蓝色、黄色、绿色 8 种颜色，每张图像只包含一辆汽车；Jeong 等人[4]建立的车辆颜色数据集，含黑色、灰色、银色、白色、蓝色、红色、黄色 7 种颜色；Tilakaratna 等人[5]建立的车辆颜色数据集，含银色、白色、黑色、灰色、天蓝色、红色、蓝色、棕色、橙色、粉色、绿色、黄色和紫色 13 种颜色。以上数据集难以满足车辆颜色精细识别的需求。此外，还没有针对更详细的车辆颜色的公共基准数据集，这严重的阻碍了相关研究的发展，尤其在进行评估时，缺乏统一的评价标准。

与此同时，随着深度学习技术的快速发展，车辆颜色识别已从最初的手动设计特征描述算子加传统分类器的方法[3-6]发展到了基于深度神经网络的方法[7-11]。特别地，以 Faster-RCNN[12]、SSD[13]、YOLO-v3[14]和 Retina-Net[15]为主的神经网络算法因良好的检测性能而被广泛应用到车辆颜色识别。然而，为了得到深层次的特征，神经网络往往需要设计很深的结构，参数较多，从而易出现过拟合现象。且运行时间较慢，在网络结构上仍有优化空间。

此外，车辆颜色的数据分布天然的呈现长尾效应[16]，因此对于车辆颜色的精细识别需要更合理的网络模型。事实上，自然环境的变化、车身灰尘的堆积、车色涂层的氧化、环境光照的变化都会导致在对车辆图像进行颜色分类时发生错误，从而影响识别准确度。

为了解决以上挑战和问题，本文自建了更大规模、颜色多样化的车辆颜色基准数据集，初步分析数据集呈现长尾效应。首先对数据进行去雾、光照调整等预处理操作。由于颜色属于底层特征，识别颜色的网络不需要太深的结构，而且车辆识别需要边缘信息，所以本文在 Faster-RCNN 网络结构的激发下构建了新的网络。通过改进后的残差网络提取颜色和边缘特征，利用多尺度特征融合层提取边缘信息，使用改进后的 Smooth L1 损失函数来消除长尾问题的影响，提出基于多层特征表示的光滑调制神经网络框架（SMNN-MFR）。最后通过可视化和消融实验也证明了我们网络设置的合理性和各模块的有效性。通过大量的实验证实，与当前最新方法相比，SMNN-MFR 在车辆 24 色识别上取得了显著的改进，识别精度达到 94.96%，比传统 Faster-RCNN 网络提高 33.47%，在 8 类车辆颜色上的识别精度为 97.25%，比同类车辆颜色识别的算法识别精度都有所提高。

本文的主要贡献可概括为：（1）构建和标注了 24 色车辆基准数据库，为更精细的车辆颜色识别提供了数据基础；（2）提出了一种基于多层特征表示的光滑调制神经网络算法，用于车辆 24 色识别；（3）为证明我们方法的有效性和优越性，我们进行了大量的扩展性实验，包括对网络特征进行可视化、对各个模块进行消融、以及多网络和多数据集的对

比。

本文其余部分组织如下：第 2 节介绍了与车辆颜色识别有关的研究工作，第 3 节详细介绍本文给出的 24 色车辆基准数据库，第 4 节详细介绍本文构造的一个基于多层特征表示的光滑调制神经网络结构，第 5 节为实验设计与结果分析，第 6 节总结了全文的研究工作。

## 2 相关工作

车辆颜色识别广泛的应用价值，已经吸引了计算机视觉领域的研究者，并提出了许多高效的识别方法[3-11]。一般地，当前存在的车辆颜色识别工作可以分为两大类，即基于手动特征的方法和基于深度学习的方法。

手动特征的方法通常依靠先验知识设计复杂的特征描述算子进行车辆颜色特征提取，然后利用 K 近邻（K-Nearest Neighbor, KNN）、支持向量机（Support Vector Machine, SVM）等经典的分类器进行分类，典型的工作有[3-6]，包括 Chen 等人[3]提取颜色直方图特征，利用空间金字塔匹配和特征上下文信息，并结合 SVM 来识别车辆颜色，准确率达 92.49%，但只有 8 种颜色类型；Jeong 等人[4]提出一种基于投票的车前图像均匀性块搜索的颜色分类方法，通过提取每个色块的 HSV 直方图，并对每个色块的颜色进行分类，将其与投票策略相结合来决定车辆的颜色，虽然准确率达 98.39%，但只有 7 种颜色类别；Tilakaratna 等人[5]讨论了与车辆牌照识别系统一起使用的车辆颜色识别系统的实现，该系统提供广泛的颜色分类，包括 13 种颜色，但最佳精度仅为 87.52％；Dule 等人[6]提出从不同的颜色空间中提取不同的颜色特征，并使用 KNN、SVM 和人工神经网络来研究车辆颜色的识别性能，在 16 种颜色空间组合提取特征并取得 83.5%的准确率，但只有 7 类车辆颜色的识别，且最终识别准确率较低。明显地，尽管各种有效的基于手工特征进行车辆颜色识别的工作被提出，但他们涉及的车辆颜色种类最多的也只有 13 类，并且存在识别准确率低的问题，所以离实际应用还有一定的距离。

近年来，深度学习模型已经在很多视觉任务上成功超越了传统手动特征的方法。深度学方法相比于手动特征的方法，能够降低特征提取过程的复杂性，且自适应的特征学习方式会使模型具有更强的泛化能力。因此，深度学习也被引入到了车辆颜色识别任务，典型的工作有[7-11]，包括 Hu 等人[7]利用空间金字塔与卷积神经网络相结合的方法获取图像的特征图，并使用传统分类器将其应用于车辆颜色识别，虽然最高准确率达到 94.61%，但只有 8 种颜色类型；Rachmadi 等人[8]采用端到端深度神经网络结构，通过联合优化将特征提取和分类器融合为统一的框架，虽然识别准确率达到了 94.47%，但只有 8 种颜色类型，且运行时间较慢；Li 等人[9]提出了一种基于分层微调策略的城市监控视频高精度车辆颜色识别方法，该方法结合了预训练和分层微调以获得可以适应光照条件变化的不同分类模型，虽然识别准确率最高达到 98.26%，但只有 10 种颜色类别；Fu 等人[10]利用基于残差网络的多尺度综合特征融合卷积神经网络用于车辆颜色识别，并建立了一个在实际交通场景中用于车辆颜色识别的系统，虽然识别准确率较高，但只有 8 种颜色类别；Nafzi 等人[11]提出了一种基于车辆型号和颜色分类的识别模块，可以由自动车辆监视或视频数据进行快速分析，虽然识别准确率达到 97.96%，但只有 10 种颜色类型。总体说来，深度卷积神经网络的引入，大幅度的提升了车辆颜色识别的准确率。然而由于卷积神经网络结构的复杂，参数较多，所以容易出现过拟合现象，且对算力有较高的要求，在网络结构上仍有优化空间。

针对数据集种类较少且分布不均，深度学习网络结构的复杂，易出现过拟合现象，复杂应用场景达不到相应的实时性和准确率等问题，本文提出一种基于多层特征表示的光滑调制神经网络方法，并在识别车辆 24 色上获得了较好的性能。

## 3 自建车辆 24 色基准数据集

由于现存公开的车辆数据集中所包含车辆颜色的种类较少，且单张图片中的车辆颜色单一，所以本文建立了颜色更加丰富的数据集，以提高车辆颜色精细分类的准确率。

本文的车辆 24 色基准数据集（Vehicle Color-24）详细构建情况如下：搜集来自西安市多处道路监控摄像头拍摄的视频共 100 小时，拍摄方向为车辆前方图像（或轻微的角度变化），截取图像 10091 张，分辨率为 1920×1080 像素，共有 31232 辆车。每张图像包含多色车辆，涵盖 24 种颜色类型，包括：红色、暗红色、粉红色、橙色、暗橙色、橙红色、黄色、柠檬黄、土黄色、绿色、暗绿色、

草绿色、青色、蓝色、暗蓝色、紫色、黑色、白色、银灰色、灰色、深灰色、香槟色、棕色、深棕色。

特别地，该数据集的数据样本都是真实自然场景下的车辆图片，车辆种类包括轿车、卡车和公交车等多种车辆类型，并且包括了大量在雾天和低照度等恶劣条件下的数据，所以本文的数据集更接近真实场景，也更具有挑战性。数据集中的部分车辆颜色样本如表 1 所示。

本文自建数据集的车辆颜色分布情况如图 1 所示，每种车辆颜色的数量如表 2 所示。其中白色车辆最多，有 11827 辆，占比高达 37.87%；紫色车辆最少，有 15 辆，占比 0.04%，可见其数据分布存在明显的长尾效应。

表 1 自建车辆 24 色基准数据集样本示例

| 颜色 | 示例图 | 颜色 | 示例图 | 颜色 | 示例图 |
|---|---|---|---|---|---|
| 红色<br>(red) | 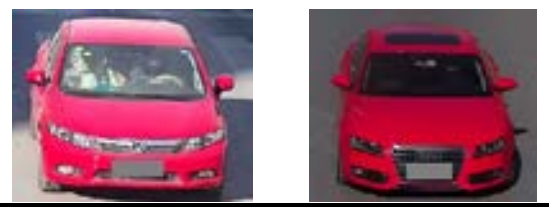 | 土黄色<br>(earthy-yellow) | 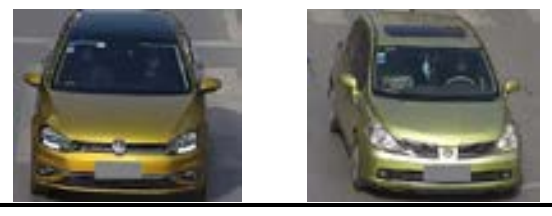 | 黑色<br>(black) | 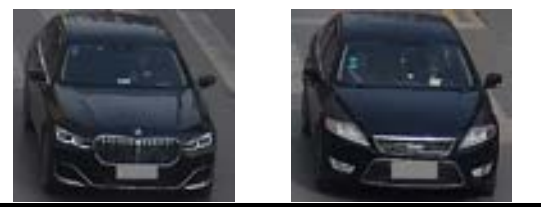 |
| 暗红色<br>(dark-red) | 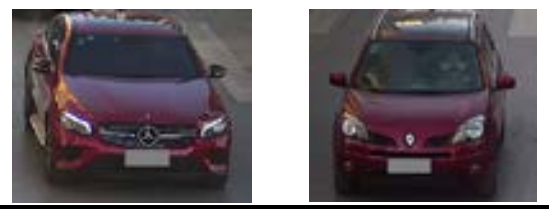 | 绿色<br>(green) | 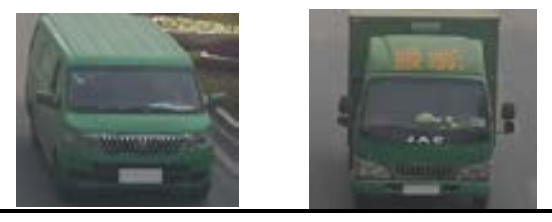 | 白色<br>(white) | 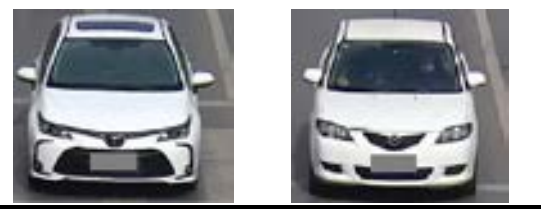 |
| 粉红色<br>(pink) | 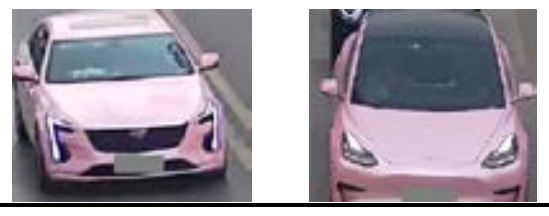 | 暗绿色<br>(dark-green) | 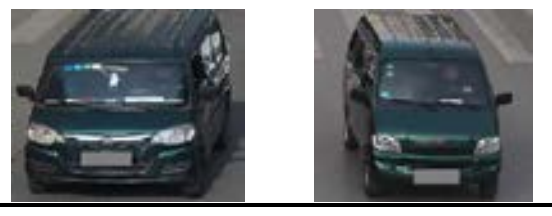 | 银灰色<br>(silver-gray) | 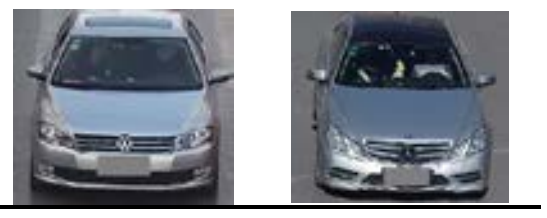 |
| 橙色<br>(orange) | 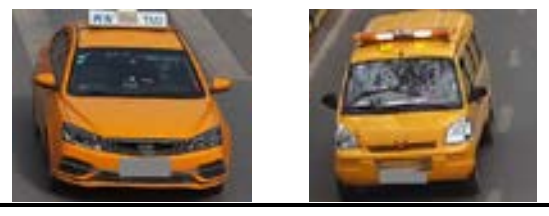 | 草绿色<br>(grass-green) | 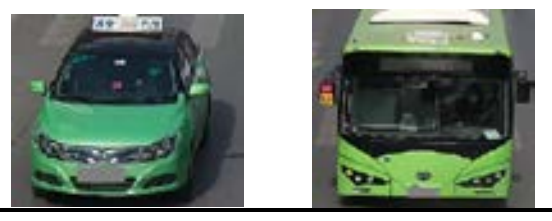 | 灰色<br>(gray) | 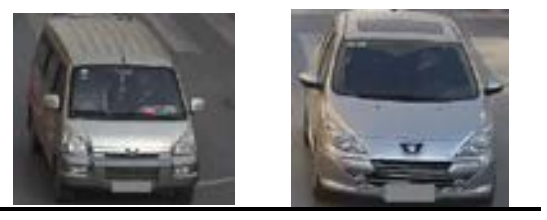 |
| 暗橙色<br>(dark-orange) | 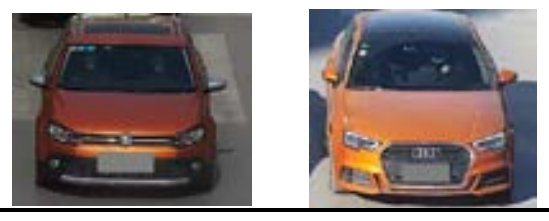 | 青色<br>(cyan) | 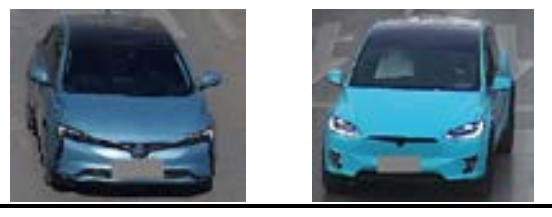 | 深灰色<br>(dark-gray) | 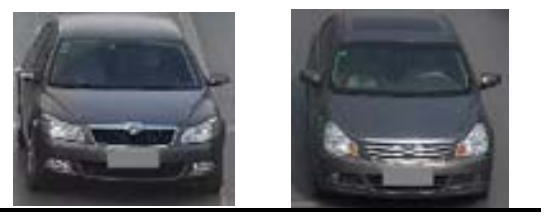 |
| 橙红色<br>(red-orange) | 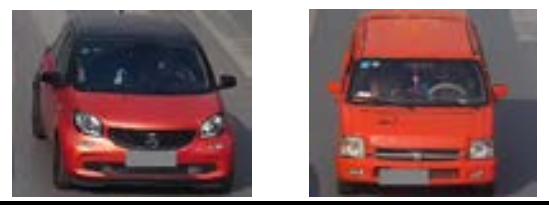 | 蓝色<br>(blue) | 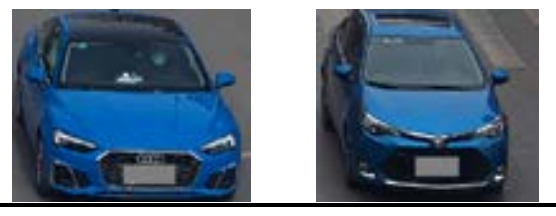 | 香槟色<br>(champagne) | 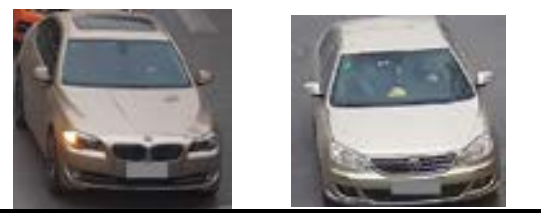 |
| 黄色<br>(yellow) | 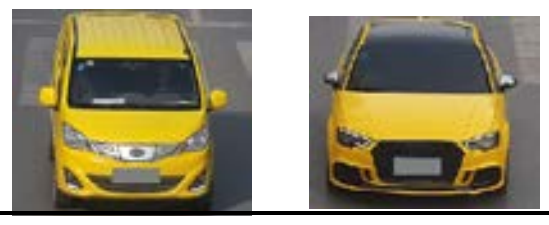 | 暗蓝色<br>(dark-blue) | 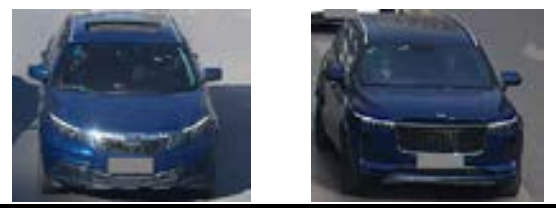 | 棕色<br>(brown) | 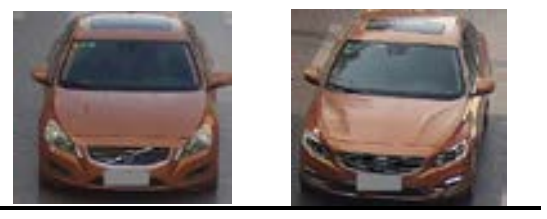 |
| 柠檬黄<br>(lemon-yellow) | 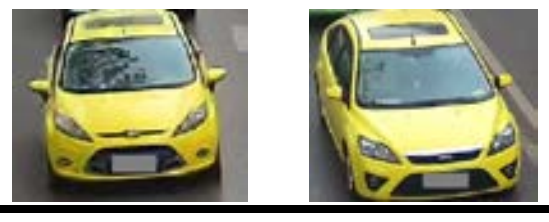 | 紫色<br>(purple) | 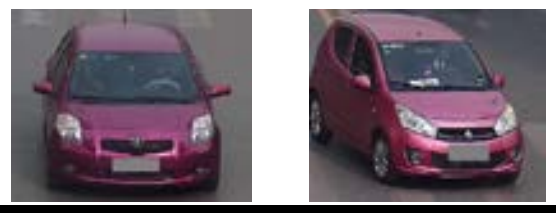 | 深棕色<br>(dark-brown) | 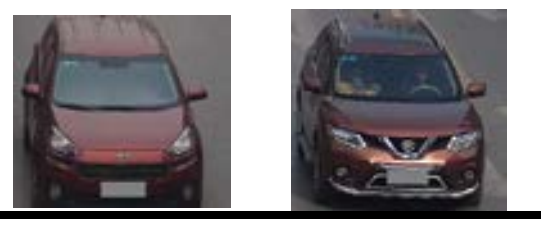 |

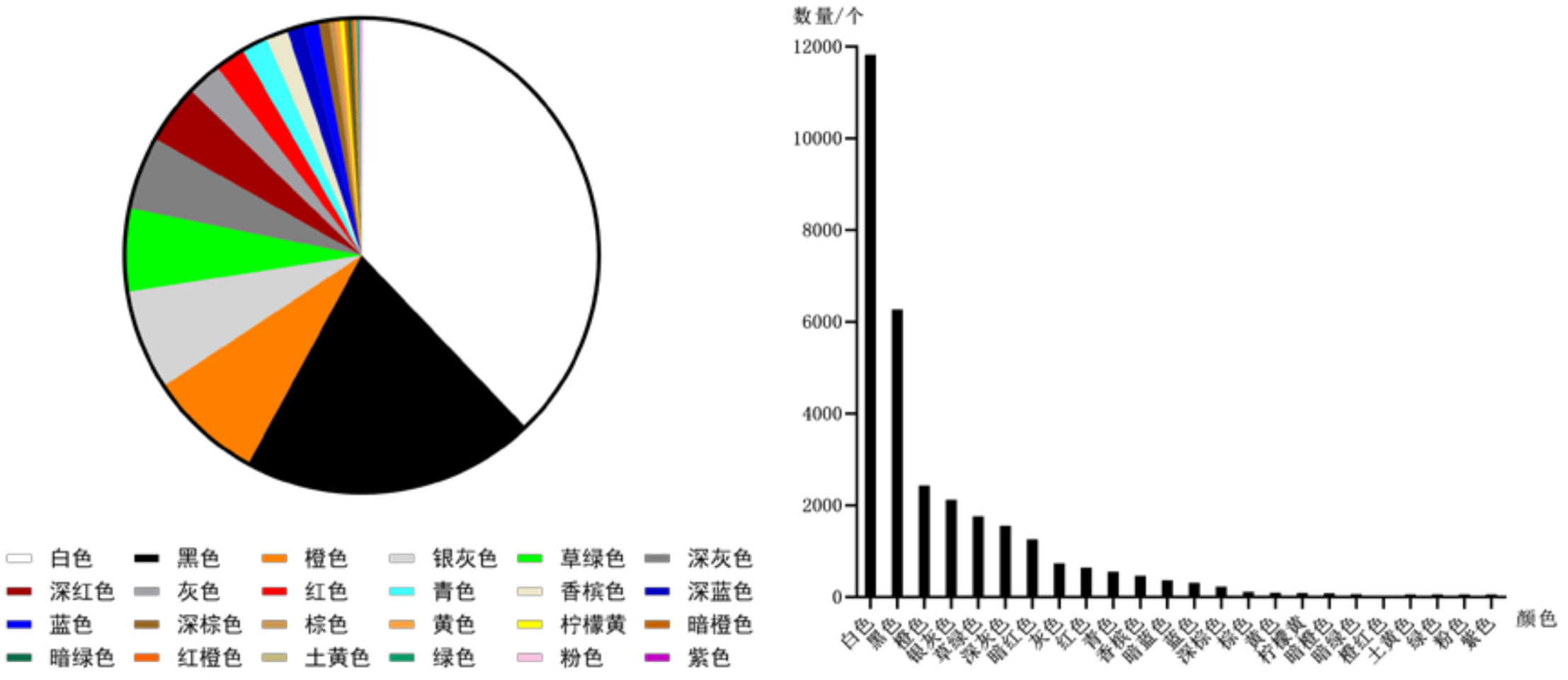

图 1 24 色车辆分布图示

表 2　24 种车辆颜色数量

| 颜色 | 数量 | 占比 | 颜色 | 数量 | 占比 | 颜色 | 数量 | 占比 |
|---|---|---|---|---|---|---|---|---|
| 白色 | 11827 | 37.87% | 红色 | 644 | 2.06% | 柠檬黄色 | 92 | 0.29% |
| 黑色 | 6270 | 20.08% | 青色 | 553 | 1.77% | 暗橙色 | 90 | 0.29% |
| 橙色 | 2431 | 7.78% | 香槟色 | 466 | 1.49% | 暗绿色 | 70 | 0.22% |
| 银灰色 | 2125 | 6.80% | 深蓝色 | 365 | 1.17% | 橙红色 | 63 | 0.20% |
| 草绿色 | 1766 | 5.65% | 蓝色 | 316 | 1.01% | 土黄色 | 52 | 0.17% |
| 深灰色 | 1555 | 4.98% | 暗棕色 | 230 | 0.74% | 绿色 | 50 | 0.16% |
| 暗红色 | 1263 | 4.04% | 棕色 | 118 | 0.38% | 粉色 | 35 | 0.11% |
| 灰色 | 736 | 2.36% | 黄色 | 100 | 0.32% | 紫色 | 15 | 0.04% |

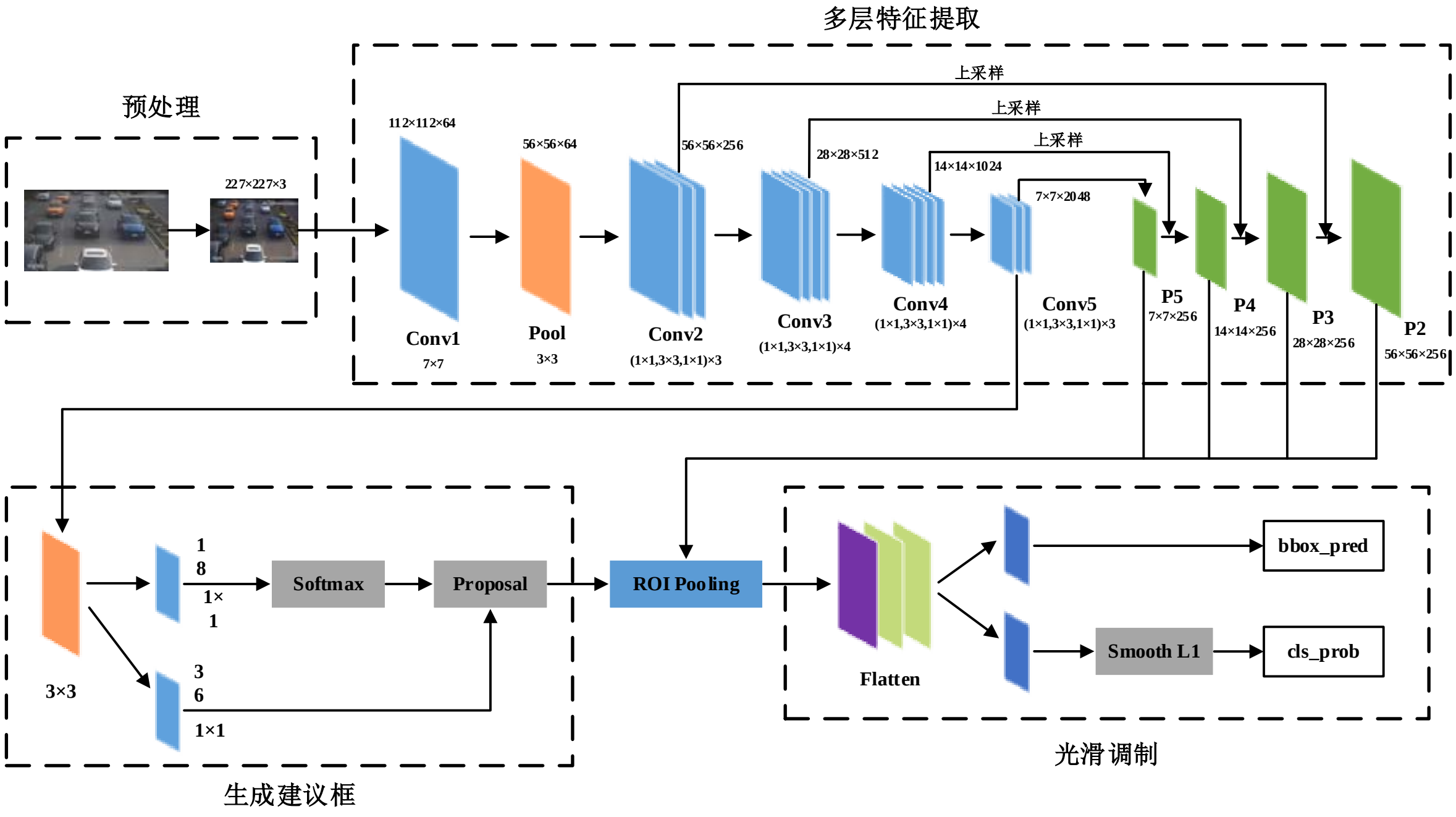

图 2　SMNN-MFR 总体结构图

# 4 基于多层特征表示的光滑调制神经网络方法（SMNN-MFR）

## 4.1 总体结构

本文的 SMNN-MFR 总体结构如图 2 所示，包括四个关键模块：预处理、多层特征提取、生成建议框、光滑调制。

（1）预处理：因为我们的数据集样本是面向自然场景收集的，所以样本图片存在雾化、低照度和过曝光等问题，所以我们首先对数据进行去雾和照度调整；

（2）多层特征提取：为了从图片中提取颜色、车型的重要特征，本文对特征提取网络进行轻量化设计，并结合多尺度特征融合网络，得到具有关键属性的特征图；

（3）生成建议框：为了获得车辆位置信息，本文在特征图上生成多个候选感兴趣区域（Region of Interest, ROI），然后利用分类器将这些 ROI 区分为背景和前景，同时利用回归器对这些 ROI 的位置进行初步的筛选；

（4）光滑调制：其中分类损失用于区分不同种类的目标，回归损失用来对目标框进行精确调整。

## 4.2 预处理

针对数据集中部分图片存在的雾化、低照度和过曝光的问题，本文首先对其进行了去雾和照度调整的操作[17]。

### 4.2.1 图像去雾算法

对于任意的输入图像，其暗通道可以用式(1)

表示：

$$J^{dark}(x)=\min_{c\in\{r,g,b\}}\left(\min_{y\in\Omega(x)}\left(J^{c}(y)\right)\right) \tag{1}$$

式中 $J^c$ 表示彩色图像的每个通道，$\Omega(x)$表示以像素 $x$ 为中心的一个窗口。

最终恢复无雾图像的公式如式(2)所示：

$$J(x)=\frac{I(x)-A}{max\left(t(x),t_0\right)}+A \tag{2}$$

其中，$J(x)$是要恢复的无雾图像，$I(x)$是待去雾的图像，$A$ 是全球大气光成分，$t(x)$为透射率，$t_0$ 为阈值，本文中所有实验均以 $t_0=0.1$ 为标准。图 3 为本文去雾前后的对比图。

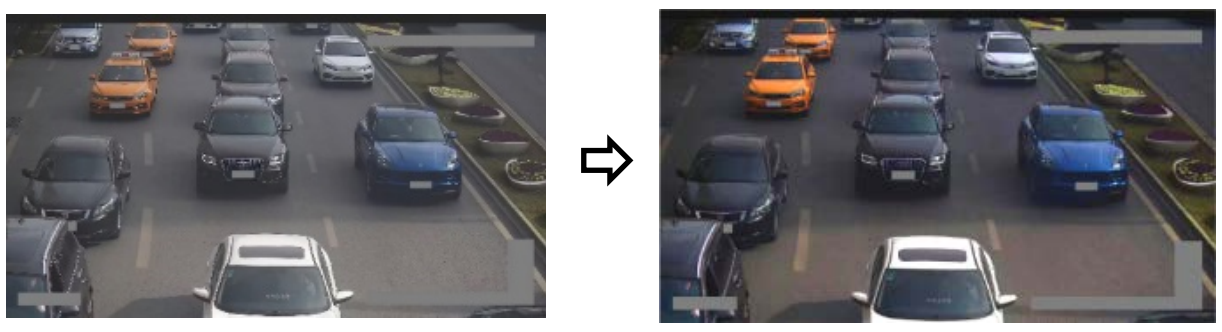

图 3 去雾前后对比图

4.2.2 照度调整算法

变化的照明会影响车辆颜色识别的性能，为了调整图像的照度，本文在 RGB 三通道上进行线性调整，如式(3)所示：

$$Val_i=\alpha\times Col_i+\beta,\ i=r,g,b \tag{3}$$

其中 $Col_i$ 代表原始图像的 RGB 值，$Val_i$ 表示调整后的 RGB 值，$\alpha$ 和 $\beta$ 是超参数。

夜间需增强图像照度时，将 $\alpha$ 设置为 1.5，$\beta$ 设置为 0，使不同颜色之间的特征差距将更加明显。正午需减少图像曝光时，将 $\alpha$ 设置为 0.8，$\beta$ 设置为 -10，可以减少整个图像的曝光。图 4 为本文光照调整前后的对比图。

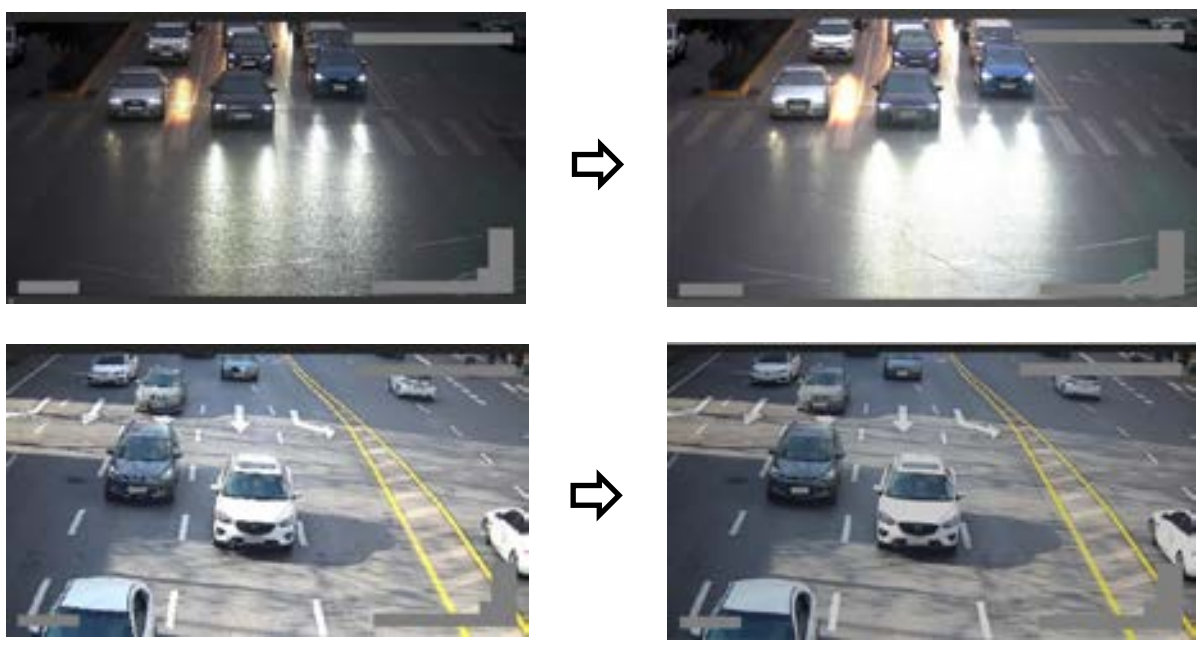

图 4 光照调整前后对比图

### 4.3 特征提取网络的确定

针对智能交通管理或刑侦判案中对嫌疑车辆追踪时频繁遭遇车牌损毁，或不易捕获到车牌的问题。而颜色作为车辆的主要特征属性，在监控视频图像中占比很大，很容易抓拍到车辆颜色属性。所以本文专注于车辆颜色特征的提取，提出 SMNN-MFR 算法，以辅助智能交通管理和刑侦判案。

事实上，颜色属于车辆的浅层特征，利用深度神经网络底层提取目标浅层特征的特点，本文参考工作[18]构造了轻量级的网络结构，并且主干网络选用网络深度及性能均表现良好的残差网络提取特征，替换原有 VGG 的特征提取网络，克服了由于网络深度加深而产生的学习效率变低，准确率无法有效提升的问题[19]。

本文利用逐层可视化的手段直观确定网络卷积模块数目，然后通过实验对比分析，给出平衡取值，在精度牺牲较小的情况下卷积模块大幅度下调，从而确定构造轻量级网络特征提取模块。本文构建的特征提取网络（Vehicle Color Recognition-ResNet，VCR-ResNet）的结构参数如表 3 所示，分为 5 个卷积块，共 42 层网络，图 5 为 VCR-ResNet 每层的可视化结果。

### 4.4 特征融合网络的确定

由于图像中存在不同尺寸的车辆目标，为了保证网络可以学习到不同尺度的颜色特征，本文引入了多尺度特征融合网络（Feature Pyramid Networks, FPN）[20]，以将深层网络和浅层网络提取的颜色特征进行合并，实现了局部特征和全局特征的融合。多尺度特征融合网络可以在速度和准确率之间进行权衡，通过它获得更加鲁棒的语义信息。由于特征像素的通道数量不同，在不同的尺度特征融合层中，数值尺度和范数也不同，在融合之前需要将每层的输出特征标准化，以使其连接不同深度的特征层。本文定义了如下的特征融合网络模块，结构框架如图 6 所示。

表 3 VCR-ResNet 的结构参数

| 操作类型 | 卷积核 | 步长 | 输出维度 | 参数数量 |
|---|---|---|---|---|
| Input | — | — | 227×227×3 | — |
| Conv1 | 7×7 | 2 | 112×112×64 | 9,408 |
| Max Pool | 3×3 | 2 | 56×56×64 | — |
| Conv2 | $\begin{bmatrix}1\times1\\3\times3\\1\times1\end{bmatrix}\times3$ | 1 | 56×56×256 | 442,368 |
| Conv3 | $\begin{bmatrix}1\times1\\3\times3\\1\times1\end{bmatrix}\times4$ | 1 | 28×28×512 | 4,718,592 |
| Conv4 | $\begin{bmatrix}1\times1\\3\times3\\1\times1\end{bmatrix}\times4$ | 1 | 14×14×1024 | 18,874,368 |
| Conv5 | $\begin{bmatrix}1\times1\\3\times3\\1\times1\end{bmatrix}\times3$ | 1 | 7×7×2048 | 56,623,104 |
| Average Pool | 7×7 | 1 | 1×1×2048 | — |
| Fc | — | — | 1000 | 2,048,000 |

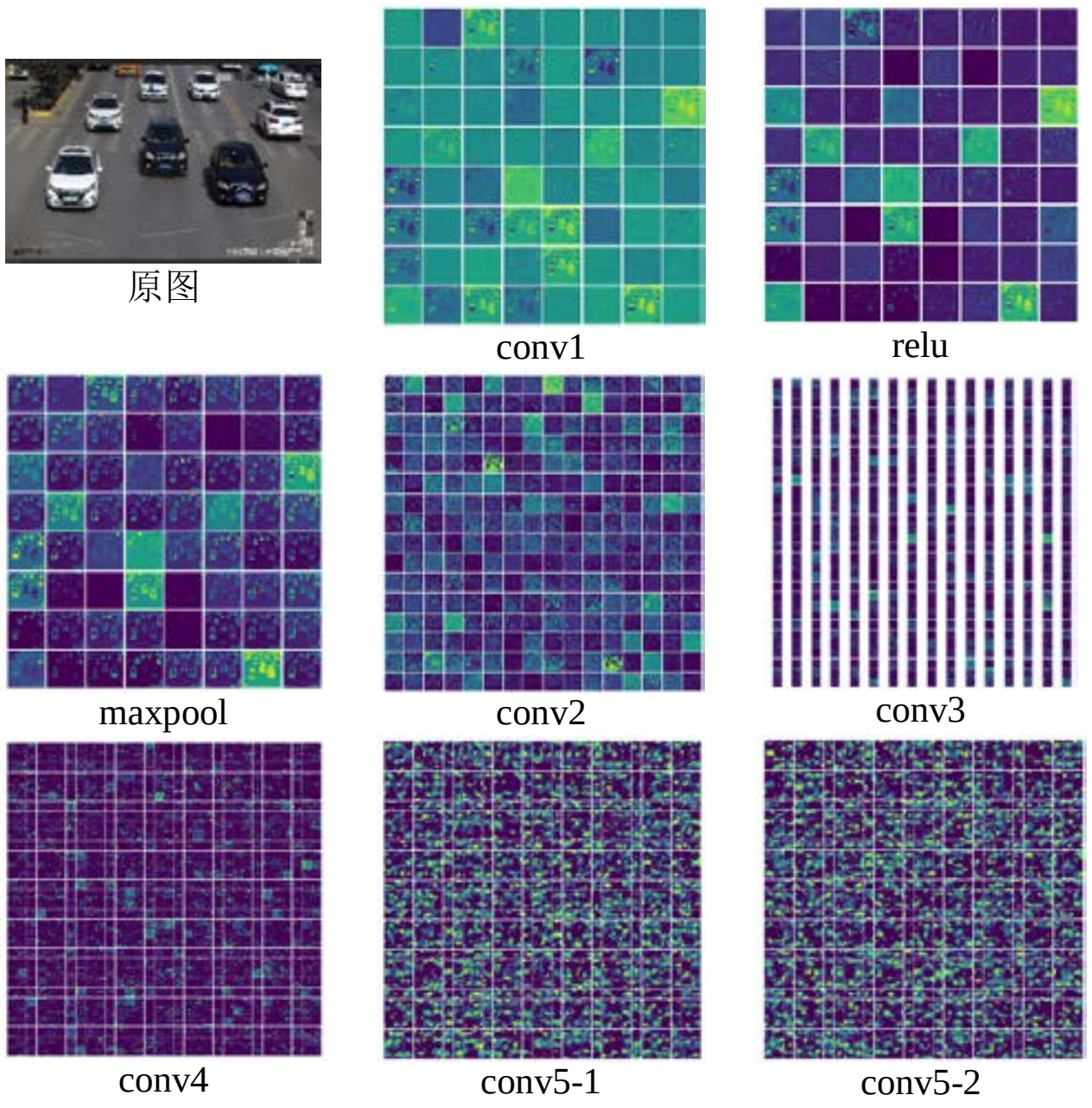

图 5　VCR-ResNet 每层可视化结果

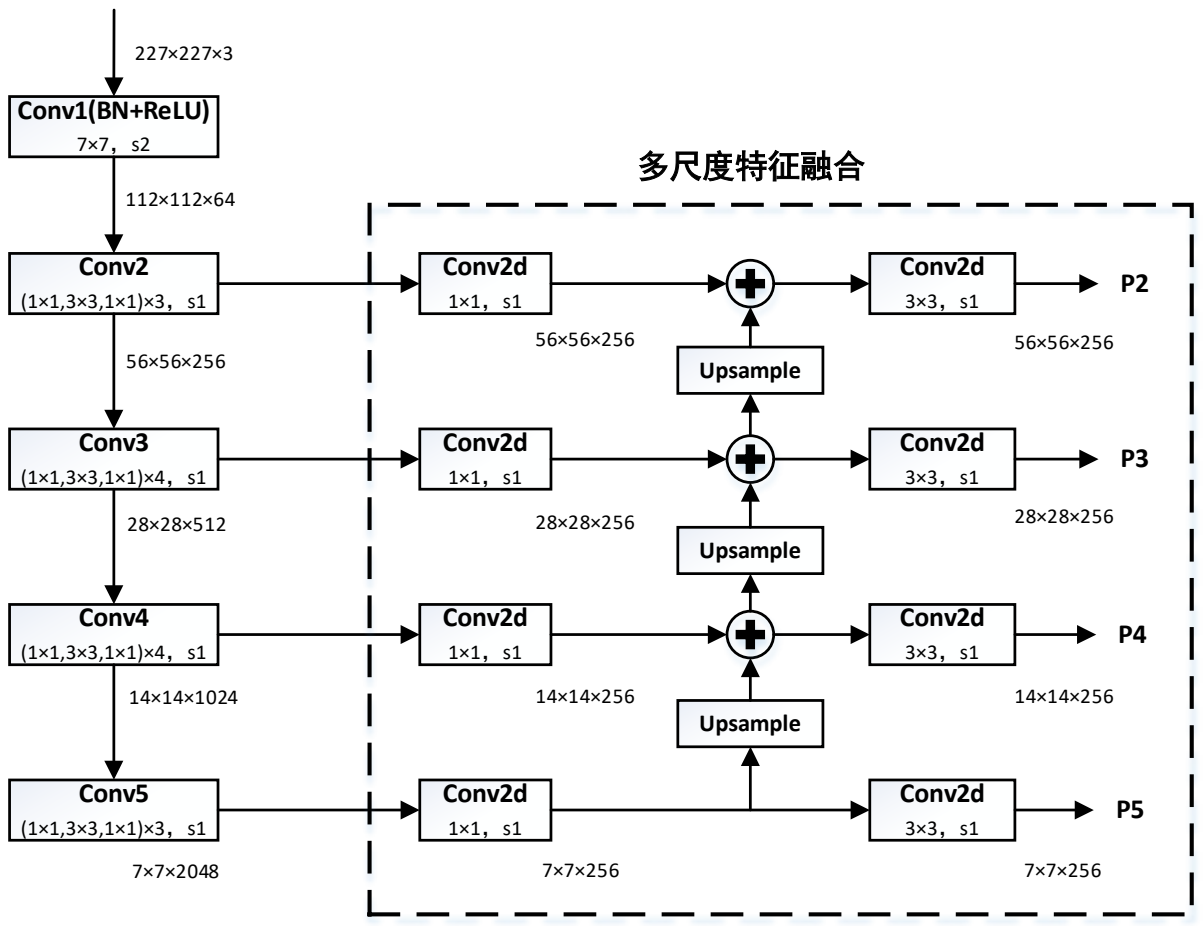

图 6　本文特征融合网络的结构图

### 4.5　损失函数的确定

损失函数实质上是计算真实输出与理想输出之间的差异，损失函数的选择直接决定了网络模型的性能。目标检测常见的损失函数包括：交叉熵损失（CE Loss）、L1 损失（L1 Loss）、均方误差损失（MSE Loss）、焦点损失（Focal loss）。实验表明，多样本类会主导模型的训练，导致模型训练退化和最终的识别准确率低。因 Smooth L1 损失函数可防止梯度爆炸，且同时拥有交叉熵损失和 L1 损失的优点，所以本文通过改进 Smooth L1 损失函数来对网络进行优化，来解决车辆颜色精细识别中存在类别不平衡的问题[21-22]。

特别地，本文通过增加超参数 $\beta$ 来优化 Smooth L1 损失，新的损失函数（Vehicle Color Recognition-Loss，VCR-Loss）如公式(4)所示：

$$\mathrm{loss}(x,y)=\frac{1}{n}\sum_{i=1}^{n}\begin{cases}0.5*\left(y_i-f\left(x_i\right)\right)^2/\beta, & \text{if } \left|y_i-f\left(x_i\right)\right|<\beta \\ \left|y_i-f\left(x_i\right)\right|-0.5*\beta, & \text{otherwise}\end{cases} \tag{4}$$

其中 $y_i$ 为真实值，$f(x_i)$为预测值，实验中 $\beta$ 的取值为 0.11。

## 5　实验和分析

### 5.1　实验环境与参数设置

本文实验算法基于深度学习框架 Pytorch 搭建，采用编程语言 Python 编写。实验配置：操作系统为 Windows10，CPU 为 Intel Core i7-8700K，内存为 16GB，GPU 为 NVIDIA GTX 1070Ti，显存为 8GB。

实验中，我们将 Vehicle Color-24 数据集按 8:1:1 的比例分为训练数据、验证数据和测试数据。在训练阶段，初始学习率设为 0.0001，batch_size 设为 4，epoch 设为 50，置信度阈值设为 0.5。

### 5.2　评价指标

本文使用平均精度均值（mean Average Precision, mAP）来衡量网络对于车辆颜色识别的性能。AP 值是由查准率（Precision）和召回率（Recall）在不同阈值下构成的 $P(R)$曲线下的面积经过插值求解计算得到的，如公式(5)所示：

$$AP=\int_0^1 P(R)dR=\sum_{k=1}^{N}\max_{k_1\geq k}P\left(k_1\right)\Delta R(k) \tag{5}$$

其中 $\max_{k_1\geq k}P\left(k_1\right)$表示对于某个召回率 $R$，所有召回率不小于 $R$ 中的查准率（Precision）最大值，$\Delta R(k)$表示召回率（Recall）的变化值，mAP 就是所有类的 AP 值的平均值。

### 5.3　实验结果分析

由于数据库呈长尾分布，本文给出多层特征表示下的光滑调制神经网络（SMNN-MFR），该网络由预处理、VCR-ResNet、FPN 和 VCR-Loss 构成。分别给出消融实验结果，给出在训练好的模型下 24 色的数据库测试结果，给出 24 色基准数据库上与其它车辆颜色检索识别算法的对比结果，以及在其它数据库中的算法的对比结果。

#### 5.3.1　消融实验分析

为了验证本文提出的方法，我们先进行预处理操作，然后在预处理的基础上通过消融实验对比经典特征提取网络 VGG12、VGG16、ResNet34、ResNet50、Mobile-Net 和 VCR-ResNet。

表 4 预处理和特征提取网络消融实验的识别精度对比表

| 对比<br>颜色 | Faster RCNN | +preprocessing | VGG13 | VGG16 | ResNet34 | ResNet50 | MobileNet | +VCR-ResNet |
|---|---|---|---|---|---|---|---|---|
| 白色 | 0.84 | 0.83 | 0.64 | 0.83 | 0.81 | 0.82 | 0.8 | 0.85 |
| 黑色 | 0.82 | 0.83 | 0.6 | 0.83 | 0.79 | 0.81 | 0.74 | 0.84 |
| 橙色 | 0.81 | 0.83 | 0.75 | 0.83 | 0.8 | 0.81 | 0.79 | 0.81 |
| 银灰色 | 0.77 | 0.78 | 0.62 | 0.78 | 0.7 | 0.71 | 0.71 | 0.73 |
| 草绿色 | 0.70 | 0.76 | 0.72 | 0.76 | 0.75 | 0.78 | 0.77 | 0.8 |
| 深灰色 | 0.66 | 0.68 | 0.57 | 0.68 | 0.63 | 0.66 | 0.62 | 0.67 |
| 暗红色 | 0.78 | 0.76 | 0.61 | 0.76 | 0.73 | 0.72 | 0.7 | 0.72 |
| 灰色 | 0.18 | 0.26 | 0.5 | 0.26 | 0.53 | 0.55 | 0.58 | 0.65 |
| 红色 | 0.60 | 0.60 | 0.58 | 0.6 | 0.7 | 0.71 | 0.7 | 0.73 |
| 青色 | 0.75 | 0.80 | 0.72 | 0.8 | 0.77 | 0.78 | 0.81 | 0.8 |
| 香槟色 | 0.63 | 0.76 | 0.59 | 0.76 | 0.72 | 0.7 | 0.67 | 0.7 |
| 深蓝色 | 0.66 | 0.75 | 0.63 | 0.75 | 0.63 | 0.68 | 0.62 | 0.7 |
| 蓝色 | 0.73 | 0.77 | 0.62 | 0.77 | 0.65 | 0.65 | 0.61 | 0.66 |
| 暗棕色 | 0.45 | 0.29 | 0.48 | 0.29 | 0.61 | 0.63 | 0.63 | 0.69 |
| 棕色 | 0.30 | 0.07 | 0.52 | 0.07 | 0.57 | 0.56 | 0.66 | 0.68 |
| 黄色 | 0.51 | 0.56 | 0.7 | 0.56 | 0.51 | 0.49 | 0.49 | 0.44 |
| 柠檬黄 | 0.87 | 0.92 | 0.7 | 0.92 | 0.55 | 0.57 | 0.56 | 0.59 |
| 暗橙色 | 0.65 | 0.52 | 0.64 | 0.52 | 0.58 | 0.57 | 0.51 | 0.62 |
| 暗绿色 | 0.38 | 0.75 | 0.63 | 0.75 | 0.54 | 0.53 | 0.54 | 0.57 |
| 橙红色 | 0.24 | 0.43 | 0.58 | 0.43 | 0.49 | 0.51 | 0.5 | 0.51 |
| 土黄色 | 0.62 | 0.33 | 0.69 | 0.33 | 0.52 | 0.51 | 0.55 | 0.6 |
| 绿色 | 0.61 | 0.97 | 0.73 | 0.97 | 0.63 | 0.65 | 0.67 | 0.69 |
| 粉色 | 0.50 | 0.50 | 0.68 | 0.5 | 0.7 | 0.72 | 0.73 | 0.71 |
| 紫色 | 0.00 | 0.00 | 0.47 | 0 | 0.55 | 0.57 | 0.58 | 0.6 |
| 平均精度 | 58.59% | **61.49%** | 62.38% | 61.49% | 64.42% | 65.38% | 64.75% | **68.17%** |
| 相对提高 | 0% | **2.90%** | 3.79% | 2.90% | 5.83% | 6.79% | 6.16% | **9.58%** |

表 5 多尺度特征融合和损失函数消融实验的识别精度对比表

| 对比<br>颜色 | Faster RCNN | +FPN | L1 Loss | CE Loss | MSE Loss | Focal Loss | +VCR-Loss |
|---|---|---|---|---|---|---|---|
| 白色 | 0.84 | 0.96 | 0.96 | 0.95 | 0.99 | 0.98 | 0.98 |
| 黑色 | 0.82 | 0.94 | 0.94 | 0.9 | 0.98 | 0.97 | 0.97 |
| 橙色 | 0.81 | 0.88 | 0.88 | 0.87 | 0.97 | 0.98 | 0.98 |
| 银灰色 | 0.77 | 0.85 | 0.85 | 0.86 | 0.93 | 0.97 | 0.96 |
| 草绿色 | 0.70 | 0.81 | 0.81 | 0.83 | 0.95 | 0.98 | 0.98 |
| 深灰色 | 0.66 | 0.73 | 0.73 | 0.78 | 0.88 | 0.94 | 0.94 |
| 暗红色 | 0.78 | 0.83 | 0.83 | 0.82 | 0.87 | 0.97 | 0.98 |
| 灰色 | 0.18 | 0.69 | 0.69 | 0.76 | 0.79 | 0.82 | 0.89 |
| 红色 | 0.60 | 0.78 | 0.78 | 0.75 | 0.74 | 0.96 | 0.96 |
| 青色 | 0.75 | 0.81 | 0.81 | 0.83 | 0.84 | 0.98 | 0.97 |
| 香槟色 | 0.63 | 0.76 | 0.76 | 0.79 | 0.8 | 0.94 | 0.91 |
| 深蓝色 | 0.66 | 0.77 | 0.77 | 0.74 | 0.75 | 0.97 | 0.96 |
| 蓝色 | 0.73 | 0.77 | 0.77 | 0.78 | 0.79 | 0.97 | 0.97 |
| 暗棕色 | 0.45 | 0.79 | 0.79 | 0.77 | 0.77 | 0.88 | 0.97 |
| 棕色 | 0.30 | 0.71 | 0.71 | 0.75 | 0.78 | 0.8 | 0.88 |
| 黄色 | 0.51 | 0.30 | 0.30 | 0.79 | 0.81 | 0.95 | 0.97 |
| 柠檬黄色 | 0.87 | 0.77 | 0.77 | 0.85 | 0.87 | 0.99 | 0.99 |
| 暗橙色 | 0.65 | 0.70 | 0.70 | 0.81 | 0.48 | 0.94 | 0.96 |
| 暗绿色 | 0.38 | 0.71 | 0.71 | 0.8 | 0.89 | 0.91 | 0.94 |
| 橙红色 | 0.24 | 0.54 | 0.54 | 0.71 | 0.57 | 0.94 | 0.99 |
| 土黄色 | 0.62 | 0.71 | 0.71 | 0.73 | 0.78 | 0.92 | 0.97 |
| 绿色 | 0.61 | 0.73 | 0.73 | 0.75 | 0.79 | 0.89 | 0.93 |
| 粉色 | 0.50 | 0.66 | 0.66 | 0.63 | 0.55 | 0.9 | 0.94 |
| 紫色 | 0.00 | 0.65 | 0.65 | 0.5 | 0.57 | 0.48 | 0.80 |
| 平均精度 | 58.59% | **74.38%** | 74.38% | 78.13% | 79.75% | 91.79% | **94.96%** |
| 相对提高 | 0% | **15.79%** | 15.79% | 19.54% | 21.16% | 33.20% | **36.37%** |

通过表 4 可以看出，经预处理后平均检测精度达到 61.49%，相对提高 2.9%，通过对比发现本文的特征提取网络框架（VCR-ResNet）更好，平均检测精度达到 68.17%，相对提高 9.58%。

然后在以上基础上继续增加多尺度特征融合层，最后在多尺度特征融合的基础上通过消融实验

对比经典损失函数 CE Loss、L1 Loss、MSE Loss、Focal Loss 和 VCR-Loss。

通过表 5 可以看出，经多尺度特征融合后平均检测精度达到 74.38%，相对提高 15.79%，通过对比发现本文的损失函数（VCR-Loss）更好，平均检测精度达到 94.96%，相对提高 36.37%。

不同模型训练损失值的对比如图 7 所示，可以看出使用本文的特征提取网络框架（VCR-ResNet）后模型训练损失值明显下降，说明模型性能有所提升。

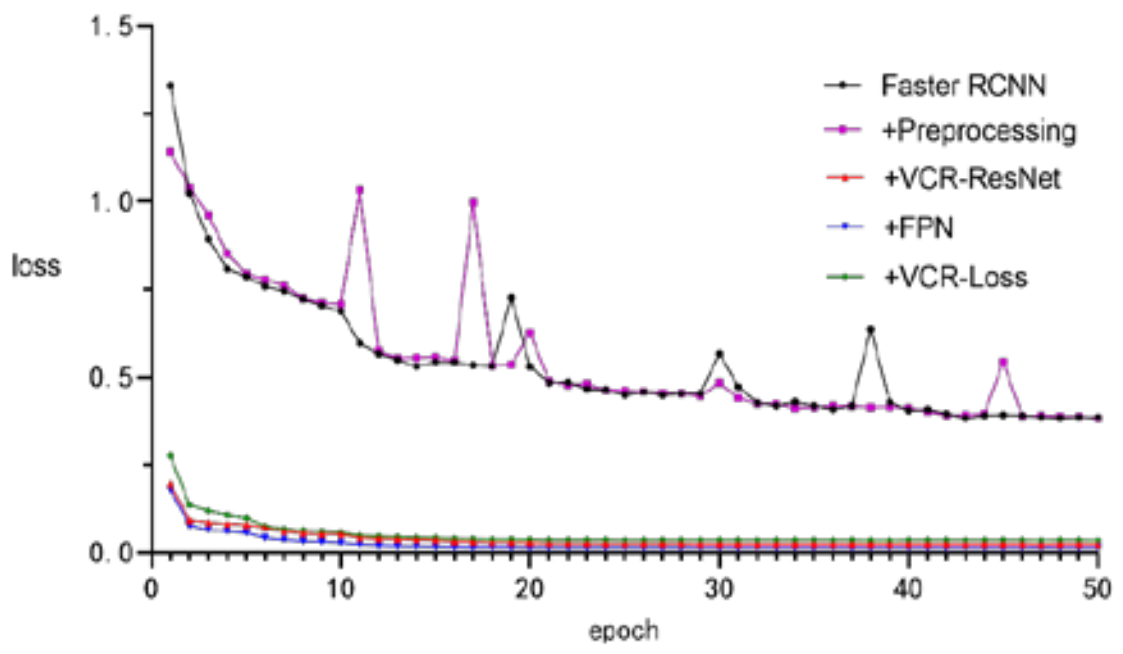

图 7　不同模型训练损失值的对比图

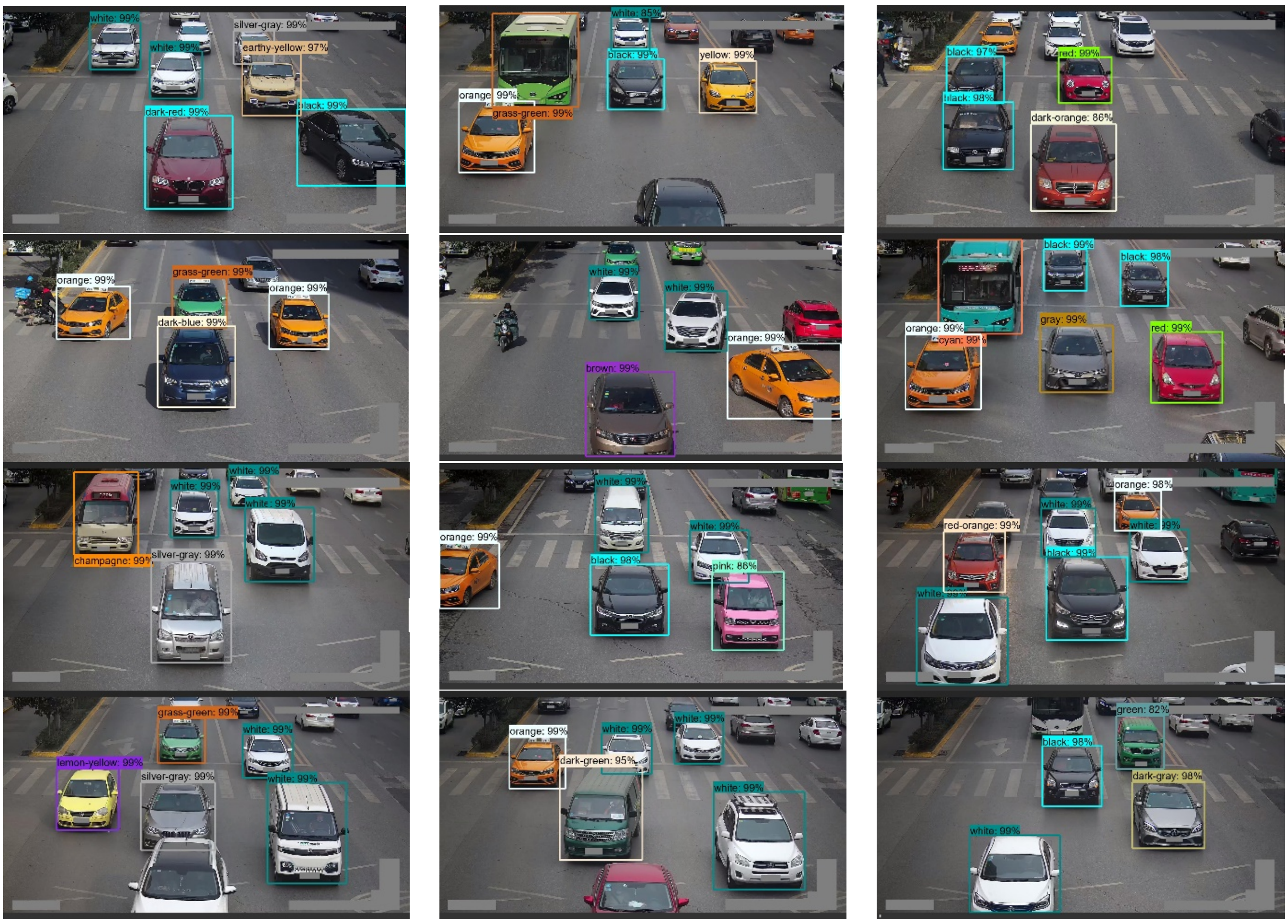

图 8　车辆 24 色识别结果

5.3.2 24 色识别结果

图 8 为 24 色识别结果图，可以看出在识别白色、黑色等样本多的类别时，识别精度较高，在识别粉色、暗橙色等样本少的类别时，识别精度相对较低。

5.3.3　不同网络对比分析

为了验证本文提出方法的有效性，我们将模型的颜色识别结果与几种经典神经网络模型的颜色识别结果进行对比，包括 Faster-RCNN[12]、SSD[13]、YOLO-v3[14]、Retina-Net[15]等。表 6 列出了以上网络在识别 24 种颜色上的准确率以及平均识别准确率。可以看出，模型在 24 类别的颜色识别中皆表现良好，本文方法的平均识别精度为 94.96%，比 Faster-RCNN 提高 33.47%，比 SSD 方法提高 16.83％，比 YOLO-v3 方法提高 18.58％，比 Retina-Net 方法提高 3.27%。

我们还将模型的识别结果与目前已提出的 8 色车辆识别进行对比，包括 Chen[3]、Hu[7]、Fu[10]等人提出的算法。表 7 列出了几种网络在识别 8 种颜色上的准确率以及平均识别准确率。可以看出，模型在 8 类别的颜色识别中皆表现良好，本文方法的平均识别精度为 97.25%，比 Chen 提出的算法提高 4.62%，比 Hu 提出的算法提高 2.87％，比 Fu 提出的算法提高 0.12%。

5.3.4 不同数据库对比分析

为了验证本文提出的方法，我们还在 Chen[3]、Jeong[4]、Tilakaratna[5]等人建立的数据库上进行了对比分析。表 8-10 列出了本文模型在以上数据库中每种颜色的识别准确率以及平均识别准确率。可以看出，本文模型的平均识别精度皆表现良好，说明本文的算法具有良好的泛化性能。

表 6 不同网络识别 24 色的精度对比表

| 颜色＼算法 | **Faster-RCNN**[12] | **SSD**[13] | **YOLO-v3**[14] | **Retina-Net**[15] | **SMNN-MFR** |
|---|---|---|---|---|---|
| 白色 | 0.84 | 0.96 | 0.97 | 0.98 | 0.98 |
| 黑色 | 0.82 | 0.95 | 0.96 | 0.97 | 0.97 |
| 橙色 | 0.81 | 0.96 | 0.97 | 0.98 | 0.98 |
| 银灰色 | 0.77 | 0.91 | 0.92 | 0.97 | 0.96 |
| 草绿色 | 0.70 | 0.96 | 0.97 | 0.98 | 0.98 |
| 深灰色 | 0.66 | 0.84 | 0.82 | 0.94 | 0.94 |
| 暗红色 | 0.78 | 0.93 | 0.94 | 0.97 | 0.98 |
| 灰色 | 0.18 | 0.54 | 0.50 | 0.82 | 0.89 |
| 红色 | 0.60 | 0.88 | 0.88 | 0.96 | 0.96 |
| 青色 | 0.75 | 0.92 | 0.93 | 0.98 | 0.97 |
| 香槟色 | 0.63 | 0.81 | 0.83 | 0.94 | 0.91 |
| 深蓝色 | 0.66 | 0.86 | 0.85 | 0.97 | 0.96 |
| 蓝色 | 0.73 | 0.87 | 0.85 | 0.97 | 0.97 |
| 暗棕色 | 0.45 | 0.71 | 0.68 | 0.88 | 0.97 |
| 棕色 | 0.30 | 0.58 | 0.52 | 0.8 | 0.88 |
| 黄色 | 0.51 | 0.79 | 0.72 | 0.95 | 0.97 |
| 柠檬黄色 | 0.87 | 0.93 | 0.84 | 0.99 | 0.99 |
| 暗橙色 | 0.65 | 0.78 | 0.66 | 0.94 | 0.96 |
| 暗绿色 | 0.38 | 0.58 | 0.63 | 0.91 | 0.94 |
| 橙红色 | 0.24 | 0.61 | 0.61 | 0.94 | 0.99 |
| 土黄色 | 0.62 | 0.74 | 0.69 | 0.92 | 0.97 |
| 绿色 | 0.61 | 0.74 | 0.77 | 0.89 | 0.93 |
| 粉色 | 0.50 | 0.71 | 0.75 | 0.9 | 0.94 |
| 紫色 | 0.00 | 0.19 | 0.08 | 0.48 | 0.80 |
| 平均精度 | 58.59% | 78.13% | 76.38% | 91.79% | **94.96%** |

表 7 不同网络识别 8 色的精度对比表

| 颜色＼算法 | **Chen**[3] | **Hu**[7] | **Fu**[10] | **SMNN-MFR** |
|---|---|---|---|---|
| 白色 | 0.94 | 0.96 | 0.98 | 0.99 |
| 黑色 | 0.97 | 0.97 | 0.99 | 0.98 |
| 灰色 | 0.85 | 0.87 | 0.96 | 0.96 |
| 红色 | 0.99 | 0.99 | 0.99 | 0.98 |
| 青色 | 0.98 | 0.99 | 0.96 | 0.98 |
| 蓝色 | 0.95 | 0.97 | 0.94 | 0.97 |
| 黄色 | 0.95 | 0.97 | 0.98 | 0.96 |
| 绿色 | 0.78 | 0.83 | 0.97 | 0.96 |
| 平均精度 | 92.63% | 94.38% | 97.13% | **97.25%** |

表 8 在 Chen 提出的数据库上对比结果

| 颜色 | 白色 | 黑色 | 灰色 | 红色 | 青色 | 蓝色 | 黄色 | 绿色 | 平均精度 |
|---|---|---|---|---|---|---|---|---|---|
| 原算法 | 0.94 | 0.97 | 0.85 | 0.99 | 0.98 | 0.95 | 0.95 | 0.78 | 92.63% |
| 本文算法 | 0.98 | 0.97 | 0.96 | 0.96 | 0.97 | 0.97 | 0.97 | 0.97 | **95.75%** |

表 9 在 Jeong 提出的数据库上对比结果

| 颜色 | 黑色 | 灰色 | 银色 | 白色 | 蓝色 | 红色 | 黄色 | 平均精度 |
|---|---|---|---|---|---|---|---|---|
| 原算法 | 0.98 | 0.97 | 0.97 | 0.98 | 0.98 | 0.99 | 0.98 | 97.85% |
| 本文算法 | 0.99 | 0.98 | 0.98 | 0.99 | 0.98 | 0.98 | 0.97 | **98.14%** |

表 10 在 Tilakaratna 提出的数据库上对比结果

| 颜色 | 银色 | 白色 | 黑色 | 灰色 | 天蓝 | 红色 | 蓝色 | 棕色 | 橙色 | 粉色 | 绿色 | 黄色 | 紫色 | 平均精度 |
|---|---|---|---|---|---|---|---|---|---|---|---|---|---|---|
| 原算法 | 0.81 | 0.85 | 0.95 | 0.86 | 0.87 | 0.94 | 0.94 | 0.97 | 0.90 | 0.91 | 0.91 | 0.93 | 0.94 | 90.62% |
| 本文算法 | 0.96 | 0.98 | 0.98 | 0.91 | 0.95 | 0.96 | 0.97 | 0.89 | 0.96 | 0.9 | 0.93 | 0.95 | 0.89 | **94.07%** |

# 6 结论

本文构建了更大规模、颜色多样化的车辆颜色 24 分类数据库，提出了一种基于多层特征表示的光滑调制神经网络下车辆 24 色长尾识别的方法，包括特征提取网络的重构、多尺度特征层的融合和损失函数的调整。实验结果表明，本文的方法在识别 24 类颜色的平均精度为 94.96%，识别 8 类颜色的平均精度为 97.25%，我们进一步提升了平均检测精度，更好的满足了车辆颜色精细分类的需要。然而，由于实际环境会受不可控因素的影响，且车辆颜色分布存在长尾效应，车辆颜色精细识别仍然有进一步的提升空间，我们将在未来的工作中进一步提高车辆颜色识别的精确度。

## 参 考 文 献

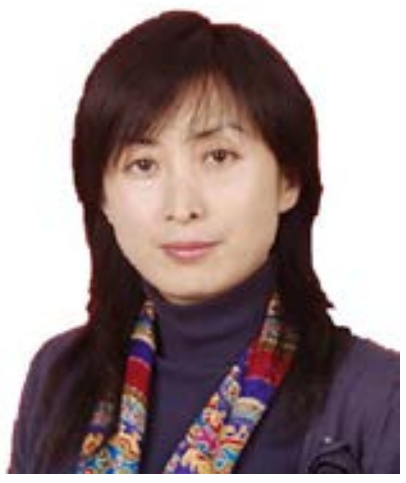

**HU Ming-Di,** PH.D., associate professor. Her research interests include image recognition, target retrieval and classification, machine learning, artificial intelligence and fuzzy information processing.

**BAI Long**, M.S. His research interests include machine learning, deep neural network and image target recognition.

**LI Ying**, M.S. Her research interests include machine learning, deep neural network and image target recognition.

**ZHAO Si-Rui**, PH.D. His research interests include computer vision, deep learning and few-shot learning.

**CHEN En-Hong**, PH.D., professor. Her research interests include machine learning, data mining and social network.

**Background**

Due to the widespread problems of poor management quality and low efficiency in intelligent transportation systems, there are more and more criminal investigation cases using vehicles as criminal tools. As another important attribute for obtaining vehicle information, vehicle color can assist traffic law enforcement and improve the reliability of vehicle identification. This research proposes a smooth modulation neural network based on multi-layer feature representation for vehicle 24-color long tail recognition method, involving machine learning image processing and target detection technology, which can improve the intelligent traffic management and criminal investigation auxiliary judgment The problem of vehicle color recognition.

Due to the complexity of paint and toning, the color of the vehicle is rich and individual. Existing vehicle color recognition research involves at most 13 types of vehicle color recognition, which is difficult to meet the needs of fine vehicle color recognition. However, deep convolutional neural networks often need to design a very deep structure, with more parameters and slower running time, so there is still room for optimization in the network structure. In addition, changes in the natural environment, accumulation of body dust, oxidation of car color coatings, and long tail effects in vehicle color data distribution will also affect the accuracy of recognition.

In order to solve the above challenges and problems, this paper builds a larger-scale, diversified vehicle 24-color benchmark data set, and then performs pre-processing operations such as defogging and lighting adjustment. Since color belongs to the underlying feature and recognition requires edge information, this paper improves the Faster-RCNN network architecture by reconstructing the feature extraction network, adding multi-scale feature fusion layers, and adjusting the loss function, and proposes a smooth modulation neural network framework based on multi-layer feature representation. Finally, ablation experiments and comparative experiments have also proved the rationality of our network settings and the effectiveness of each module.

This work was supported by the Key Research and Development Project of Shaanxi Province of China (Grant no. 2018KW-050).